LETTER

# Self-Organised Fractional Quantisation in a Hole Quantum Wire

Y. Gul[1*], S. N. Holmes[2†], M. Myronov[3], S. Kumar[1] and M. Pepper[1]

[1] London Centre for Nanotechnology and Department of Electronic and Electrical Engineering, University College London, Torrington Place, London, WC1E 7JE, United Kingdom

[2] Toshiba Research Europe Limited, Cambridge Research Laboratory, 208 Cambridge Science Park, Milton Road, CB4 0GZ, United Kingdom

[3] Department of Physics, University of Warwick, Coventry, CV4 7AL, United Kingdom

**Abstract**

We have investigated hole transport in quantum wires formed by electrostatic confinement in strained germanium two-dimensional layers. The ballistic conductance characteristics show the regular staircase of quantum levels with plateaux at $n2e^2/h$, where n is an integer, e is the fundamental unit of charge and h is Planck's constant. However as the carrier concentration is reduced, the quantised levels show a behaviour that is indicative of the formation of a zig-zag structure and new quantised plateaux appear at low temperatures. In units of $2e^2/h$ the new quantised levels correspond to values of n = 1/4 reducing to 1/8 in the presence of a strong parallel magnetic field which lifts the spin degeneracy but does not quantise the wavefunction. A further plateau is observed corresponding to n = 1/32 which does not change in the presence of a parallel magnetic field. These values indicate that the system is behaving as if charge was fractionalised with values e/2 and e/4, possible mechanisms are discussed.

\* corresponding author email: y.gul@ucl.ac.uk
† submitting author email: s.holmes@crl.toshiba.co.uk

ORCID id
S. N. Holmes: https://orcid.org/0000-0003-2776-357x
M. Pepper: https://orcid.org/0000-0003-3052-5425

# 1. Introduction to p-type Ge

Studies of quasi-one dimensional (1D) quantised conduction have been largely confined to group III-V semiconductors and silicon. In the work described here we have investigated quasi-1D transport in the ballistic regime in compressively strained germanium (s-Ge) epilayers grown on standard Si (001) substrates via an intermediate relaxed $Si_{1-x}Ge_x$ buffer. Detailed growth information has been reported elsewhere [1]. This strain results in a much lighter effective mass and a pinning of the total angular momentum (J) axis normal to the two-dimensional (2D) hole gas. The hole states differ from the electrons in that the basis states are p orbitals and so have orbital angular momentum (L) = 1. The values of J are 3/2 and 1/2. As there is spin-orbit (S-O) coupling in the material so there is an energy gap between these two states. In bulk Ge the gap is sufficiently large (0.29 eV) that we only need to consider the J = 3/2 state which is four-fold degenerate. The states are degenerate at linear momentum (k) = 0 in bulk Ge but in 2D the quantizing electric field which is normal to the plane lifts the degeneracy and a subband is formed with Heavy Hole (HH) states normal to the plane, this level forms the ground state with a Light Hole (LH) subband separated from the HH band by a gap ($\Delta$). The HHs in strained Ge have a low effective mass (m*) 0.035-0.063 $m_e$ [2] where $m_e$ is the free electron mass, compared to the heavier holes mass in GaAs [3]. Previous work [4] identified a weak S-O coupling, due to Rashba in this system with an odd-filling factor dominant magnetoresistance [5].

This letter is divided into five further sections. In section 2 one-dimensional confinement using the split-gate technique is introduced. Section 3 is a summary of the experimental details. In section 4 the experimental results on conductivity measurements are presented then discussed in section 5. A comparison is made to the stable zig-zag charge configuration. Section 6 is a final summary of the salient points consistent with fractional quantisation.

# 2. Introduction to One-Dimensional Confinement

Variable electrostatic confinement using split-gates has proved to be an effective method for converting a 2D electron gas into 1D [6]. When the channel is long, quantised magneto-electric subbands reveal the size quantisation leading to the one-dimensionality [7]. When the channel is sufficiently short, such that transport is ballistic, conductance quantisation is found at low temperatures. This has been extensively studied for electrons in GaAs [8, 9]. In addition to the quantisation of conductance (G) at $n(2e^2/h)$, where n is an integer, there is a subsidiary non-integer plateau at $0.7(2e^2/h)$. Both theoretical and experimental work [10] has shown the 0.7 structure is due to a spin separation leading to a polarisation. The establishment of backscattering, by shaping the gates imposing confinement leads to observation of a pronounced Kondo peak superimposed on the 0.7 feature showing the difference in origin of these features and eliminating the possibility that the 0.7 structure was caused by interaction with a bound state [11]. Measurement of the ac and dc conductance as a function of source-drain voltage ($V_{sd}$) allows measurement of the 1D subband structure [12] and hence Landé g-factor (in applied magnetic fields) as well as observation of half-integer plateaux, $(n+1/2)2e^2/h$ in the differential conductance ($dI/dV_{sd}$) where I is the current. At split-gate voltages ($V_{sg}$) such that G is $< 2e^2/h$, the differential conductance decreases with increasing $V_{sd}$ and reaches a plateau value of $0.25(2e^2/h)$. This value of differential conductance suggests that the momentum degeneracy is lifted and the current is spin polarised as has been demonstrated by focussing [13].

There has been considerable theoretical and experimental work on the role of the electron-electron interaction in determining energy levels in a quantum wire. Theory suggests that as the confinement weakens, or the electron-electron repulsion increases, a line of electrons splits to form a zig-zag array [14-17]. This is predicted to occur when $r_o = (2e^2/\varepsilon m^*\omega^2)^{1/3}$ where $r_o$ is the distance at which the Coulomb repulsion between a row of holes is equal to the lateral harmonic confinement energy ($\hbar\omega$), $\varepsilon$ is the dielectric constant. Increasing the 1D carrier concentration ($n_{1d}$) increases the strength of the interaction provided that the Fermi energy is not significant. Calculations show that for values of $\nu < 0.78$ a 1D line of charge is stable, where $\nu$ is a dimensionless charge separation parameter, $\nu = r_o n_{1d}$. The range of 1D carrier concentrations used in this work is such that when $1/n_{1d} < a_B$ where $a_B$ is the Bohr radius, a 1D Wigner crystal will form. In the Ge HH valence band the Bohr radius is ~ 18 nm and this sets the length scale for the types of interaction. In the case of $0.78 < \nu < 1.75$ a 1D crystal is not stable and a zig-zag configuration will form due to the increase in energy of the carriers allowing movement against the confinement potential. Increasing the interaction will cause the triangular configuration of carriers to become more acute, so that the former second nearest neighbours now become nearest neighbours [14-17]. A complex spin texture has been proposed in this arrangement with an equilateral structure occurring at $\nu = 1.46$. For $\nu > 1.75$ a large number of charge rows form resulting in a glassy arrangement. The theory of spin configurations indicated complex phases forming due to the occurrence of ring exchanges involving three and four electrons, predicted for other systems, quoted in references [14-17].

Experiments have shown that as the confinement potential is relaxed, the first plateau of quantised conductance at $2e^2/h$ weakens and can disappear with the $4e^2/h$ strengthening [18-20]. The disappearance of $2e^2/h$ indicated that the ground state was now two separate rows. An analysis of the movement of the energy levels indicated that the former first excited level crossed or anti-crossed the ground state to form a new ground state. This state would have two lateral centres of charge comprising two rows consistent with formation of a zig-zag arrangement. This conclusion has been confirmed by focussing experiments which show that a single charge centre in the channel splits into two as the channel widens and confinement is relaxed [21].

## 3. Experimental methods

In this letter we report an investigation of the conductivity of holes in 1D patterned s-Ge devices that are controlled by top-gates and split-gates. Initial 2D carrier concentration and low temperature mobility were $1.3 \times 10^{11}$ cm$^{-2}$ and $0.5 \times 10^6$ cm$^2$/V.s respectively. The measurement temperature was 20 mK in a cryofree dilution refrigerator. The lithographic separation of the split-gates was 600 nm gate separation and 200 nm length. The top-gate was separated from the Ge quantum well structure by 40 nm thick $SiO_2$ and a 70 nm thick undoped $Si_{0.3}Ge_{0.7}$ barrier. Initially we investigated the quantised plateau which appear in the differential conductance $dI/dV_{sd}$ when $V_{sd} \sim 0$ so that the experimental value is the conductance that is measured as the channel is progressively narrowed with the split-gate voltage. An excitation voltage of 10 μV was used to measure $dI/dV_{sd}$ with a current pre-amplifier set to 1V/μA. The overall carrier concentration is set by the top-gate voltage ($V_{tg}$) and in this work we investigated values down to $\sim 10^{10}$ cm$^{-2}$. The smoothness of the experimental data shows that there is no scattering in the 1D channel and mesoscopic fluctuations are absent. Six devices in total were investigated and all showed the extra plateau structure reported here below $2e^2/h$. The inset of figure 1 shows a device schematic where the source (S) and drain (D) ohmic contacts are shown with the split-gates (where $V_{sg}$ is applied) and top-gate (where $V_{tg}$ can be applied).

## 4. Experimental results

Previously published work [22] on quantum wires in s-Ge showed clear integer conductance plateaux up to $10(2e^2/h)$ with 1D subband energy spacing $\sim 1$ meV. Similar conductance structure is observed in the set of devices reported here.

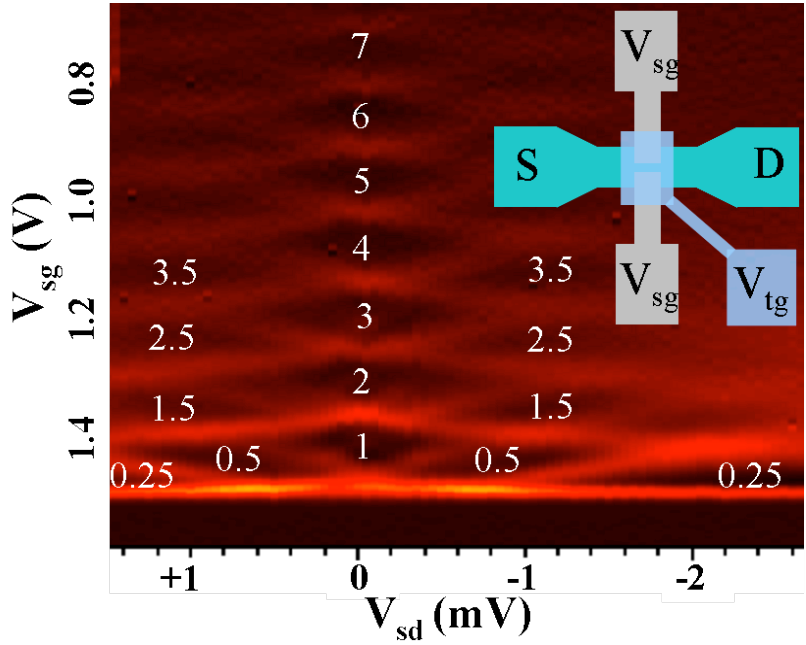

**Figure 1** The transconductance (dG/dV$_{sg}$) as a function of split-gate voltage (V$_{sg}$) on the y-axis and applied dc voltage (V$_{sd}$) on the x-axis. Integer plateaux in conductance (black areas) are labelled from depletion of the channel to 7(2e$^2$/h). Orange and red areas correspond to risers. The top-gate voltage is set to 0 V where the density is $1.3 \times 10^{11}$ cm$^{-2}$. The inset is discussed in the main text.

Application of a dc source-drain voltage splits the conductance plateaux into a series of ½ integer plateaux [12] as discussed in section 2. This is clearly seen here for the conventional 1D states shown in figure 1, where plateaux at (n+1/2)(2e$^2$/h) are visible at finite V$_{sd}$ voltage. Only the ½ integer plateaux up to 3.5 are labelled. In the transconductance the black areas correspond to conductance plateaux and the red orange regions to the risers. The transconductance is advantageous in that its behaviour as the channel is altered allows the variation in the 1D energy levels to be shown clearly.

Figure 2 shows the transconductance (dG/dV$_{sg}$) under similar experimental conditions but in the ohmic regime, i.e. V$_{sd}$ = 0 V, with a depleting voltage on the top-gate. When the top-gate becomes increasingly positive the hole concentration decreases and the confinement field weakens for a particular plateau, consequently the combination of gates allows observation of the transition from strong to weak confinement. This figure illustrates the behaviour of the transconductance as a function of both top-gate and split-gate voltages, bottom right is higher carrier concentration and strong confinement and top left is lower carrier concentration and weaker confinement. The colour plot illustrates the transconductance so that dark regions have small gradient, i.e. plateaux, and the coloured regions are the risers between the plateaux. The values of plateau conductance in units of 2e$^2$/h are at the bottom of the figure. As seen in the strong confinement region, the system behaves in the expected manner. Reading from right to left there is a conductance staircase represented by values of n(2e$^2$/h) where n is 1, 2, 3,… These correspond to the same states that are labelled in figure 1. However as the confinement is weakened, the ground state (n = 1) disappears and the former n = 2 state is now the lowest integer plateau. The disappearance of the n = 1 is accompanied by the appearance of two new states with values of n = 0.5 and 0.25 in units of $2e^2/h$.

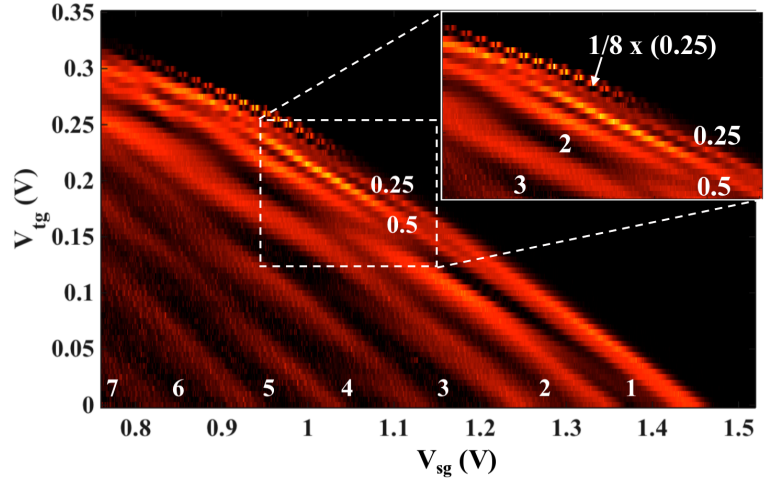

**Figure 2** The transconductance ($dG/dV_{sg}$) plotted as a function of top-gate voltage on the y-axis and split-gate voltage on the x-axis. The dc voltage is set to 0 V. Regions of orange and red correspond to large $dG/dV_{sg}$ indicating the locations of subband transitions. Black corresponds to a conductance plateau in $dG/dV_{sg}$. The inset is an expanded region around $V_{tg}$ = +0.18 V.

The 0.5 is a spin resolved state which disappears as the confinement is further weakened leaving the 0.25 with a further split off state which is more clearly shown in the conductance plot in figure 3. This figure shows that at zero magnetic field there are two states, the 0.25, (in zero magnetic field this state is nearer 0.28 and is discussed later in this letter) and a small plateau close to $(1/32)(2e^2/h)$. Increasing the magnetic field, which is parallel to the current direction in the plane of the hole gas, causes the first plateau to move closer to 0.25 and then drop rapidly before forming a new plateau close to $(0.125)(2e^2/h)$. The small 1/32 plateau increases very slightly with magnetic field to be closer to the value of $(1/32)(2e^2/h)$. As seen there is no structure up to $0.5(2e^2/h)$ as the system is in the regime of the conductance jump to $4e^2/h$. We now refer to the plateaux as the 1/4, which is spin degenerate, and 1/32 which is spin polarised as it is virtually unaffected by the magnetic field, see figure 3, both fractions are in units of $2e^2/h$.

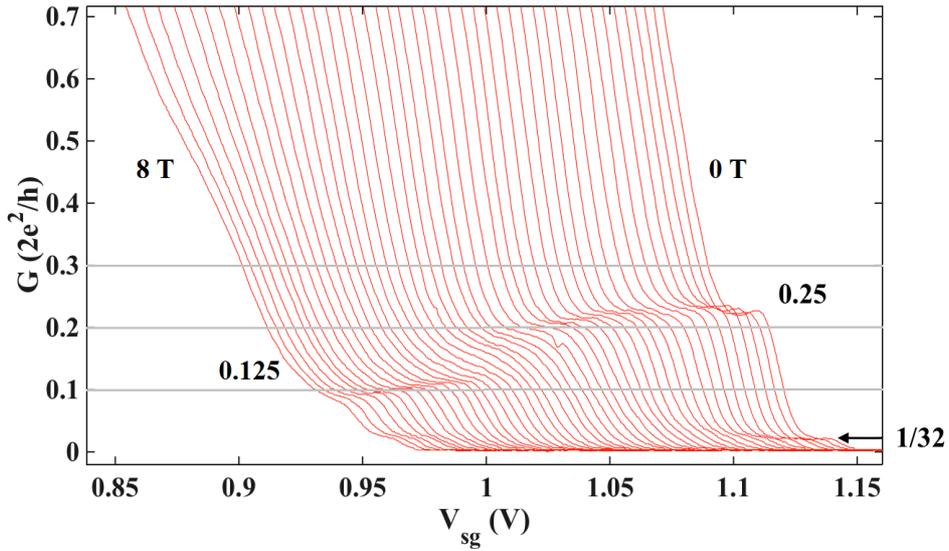

**Figure 3** The differential conductance (dI/dV$_{sd}$) as a function of split-gate voltage. V$_{tg}$ was set to +0.18 V and the dc voltage is at 0 V. The in-plane magnetic field (parallel to the current direction) was stepped by 0.2 T from 0 T to 8 T. The data has been offset horizontally for clarity.

It was also found that increasing the source-drain voltage above a magnitude of 0.25 mV caused the two new plateaux (1/4 and 1/32) to disappear and with further increase in source-drain voltage, the regular spin polarised 0.25 plateau appeared. It seems that the ¼ and 1/32 are linked as both behave in a similar manner with dc source-drain voltage. Although the 1/32 seems spin polarised, it disappears for that interval of magnetic field (~ 2 to 4 T) when the ¼ (i.e. 0.25) is gradually dropping to half its value (i.e. 1/8 or 0.125) and only reappears when the 1/8 strengthens, see figure 3. The 1/32 plateau comes up slightly higher in conductance at 8T compared to 0T.

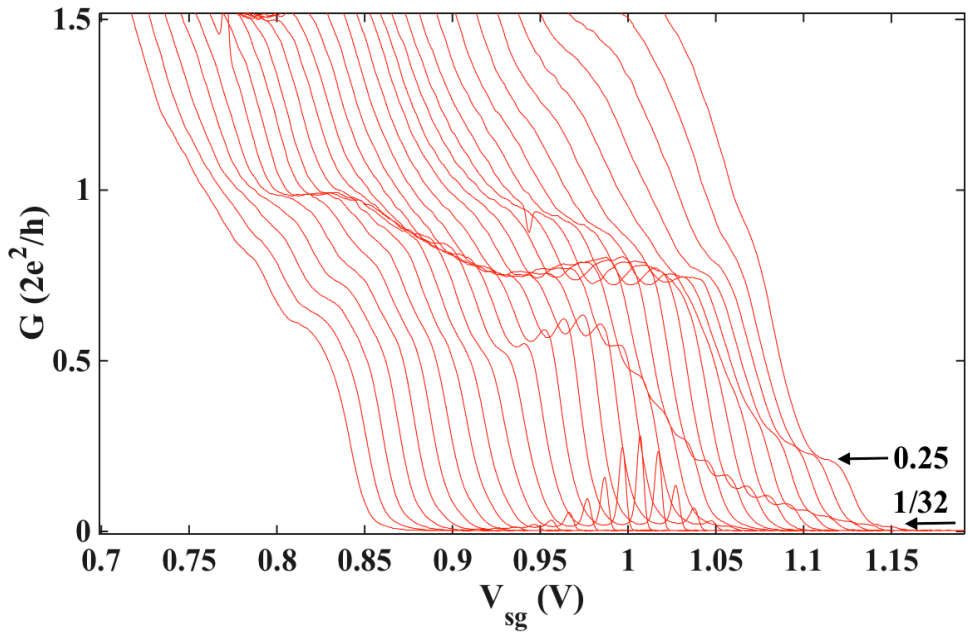

**Figure 4** The differential conductance (dI/dV$_{sd}$) as a function of asymmetric split-gate voltage (V$_{sg}$). V$_{tg}$ was set to +0.18 V with the dc voltage set to 0 V. The gate voltage was stepped from ΔV = 0 to -0.8 V in units of 20 mV.

The plateau stability was investigated by the application of an offset voltage to one of the split-gates. The results are shown in figure 4 where it is seen that the n = 1, which is at $2e^2/h$ disappears and then returns as the offset bias alters and creates a change in shape of the confining potential. Similar results have been obtained for electrons in GaAs. However the new plateaux at ¼ and 1/32 are suppressed in GaAs. The conductance with an asymmetric voltage applied to the split-gates is discussed further in section 5.

## 5. Discussion

As the system is degenerate the closeness to quantised fractional values implies that the transmission factor is close to unity and the role of the magnetic field in either halving or not affecting plateaux shows that spin incoherence is not an issue, [23]. The fractionalization is unexpected and at present there is no precise explanation, however there are clear features which we now discuss.

In the fractional quantum Hall effect (FQHE) n takes a value which is less than one. Electrons or holes behave [24, 25] as if their charge is e/3 and so n takes a value of 1/3 in units of $e^2/h$, as the spin degeneracy is lifted. Further fractional states formed at different values of carrier concentration or magnetic field if the mobility was high enough. This situation is determined by a gapped state such that gaps in Landau levels form at fractional filling factors determined by the fractional charge. If a gapped state formed in the quantum wires investigated here then this would imply that we are observing n = 1/4 with a fractional charge of e/4 in a spin degenerate configuration. The state with n = 1/32 which appears at low carrier concentration and is spin split would correspond to a fractional charge of e/32. On the other hand it has been suggested that in fractionalized topological insulators [26] if the charge is e/p then the conductance for spin degenerate levels is $(1/p^2)(2e^2/h)$, which would suggest the more reasonable fractional charges of e/2 and e/4. This conclusion is strengthened as the change in potential energy of the e/3 quasi-particles the FQHE when a voltage V is applied is given as eV/3 not eV. Deriving a quantised conductance in the normal way gives the value $(1/p^2)(2e^2/h)$ and in our experiments the fractional charges correspond to e/2 (for p = 2) and e/4 (for p = 4).

Fractional quantisation arising from fractional charge, in the absence of a quantising magnetic field, has been proposed for different experimental situations, commencing with the Su-Schrieffer-Heeger model [27, 28] of an e/2 domain wall propagation due to a defect between two degenerate phases of polyacetylene dimer. However, as the domain wall boundaries come in pairs the resultant charge is unity rather than fractional, such features in this material were not observed due to a high level of disorder. Other predictions are based around a Mott Insulator [29] or domain boundary between two topological phases in 1D [26]. It has been shown that there is a measurable 2D S-O interaction in the Ge utilised here but it is of Rashba type [4] induced by a negative gate voltage and higher carrier concentration than used in the work reported here. In our device scheme we have used a positive top-gate voltage to deplete the carriers and achieve a 2D carrier concentration of $\sim 10^{10}$ cm$^{-2}$ which is necessary for observation of the fractions.

The results reported here imply that there could be states where n has a value of 1/2 at conductance $1/4(2e^2/h)$, which is spin degenerate but persists when the degeneracy is lifted, and n = 1/4 at conductance $1/32(2e^2/h)$. This 1/4 state appears spin polarised and stays that way in the presence of a magnetic field. The new states are delicate and disappear with increasing $V_{sd}$, the 1/4 state moves up to disappear when near n = 1, i.e. a regular 1D quantum state [22, 30]. Further increase in $V_{sd}$ results in the formation of the normal spin polarised 0.25 state in differential conductance.

Previous experimental work on 1D quantum wires has established that in the regime of strong interactions and weak confinement, the ground state, comprising a single line of electrons, breaks up to form an Incipient Wigner Lattice [18-20], such a zig-zag configuration has been investigated theoretically with the prediction of many possible spin phases [14-17]. These arise from the formation of complex states initially dimerized and then ring exchanges involving three or four particles, depending on the carrier concentration and the confinement potential. Even number particle states have an odd number of permutations of exchange and support anti-ferromagnetism whereas an odd number of particles has an even number of exchange permutations and is ferromagnetic. Consequently, a dimer state will be spin degenerate whereas a three particle ring exchange will be spin polarised.

We note that the process of dimerization is enhanced in 2D if the basis functions are p orbitals which is the case for the hole system studied here. It is possible that the strong dimerization is responsible for a charge of e/2, which is spin degenerate, however such a state is unlikely to persist in the presence of a strong magnetic field and so we suggest that this is an unlikely mechanism. The 3 ring exchange is spin polarised as are the e/4 states and as the channel is widened it is possible for more complex states to form which are spin degenerate and produce the e/2, however this state must remain as the spins become parallel due to the strong magnetic field. It may be possible for two three particle rings, with opposite spin directions, to link or hybridise so giving the fractional charge and the state persists with parallel spins in the presence of the magnetic field. It is also noteworthy that if the zig-zag evolves to form three rows, as predicted in classical considerations, then it is possible to form a honeycomb configuration with many predicted exotic phases [31], with we suggest a possibility of fractional charges. At present we are not aware of a theoretical suggestion that strong ring exchange can give rise to a fractional charge, however cyclic motion in an exchange ring has similarities to the FQHE. The lighter mass in the system studied here and correspondingly large Bohr orbit increases the wavefunction overlap in the ring exchange which is further strengthened by the basis states being p orbitals, as has been stressed for interacting 2D systems [32].

Reference [33] discussed how compressing Helium 3 will reduce the duct volume linking particle positions in the exchange cycle so reducing the effects of ring exchange. In a similar manner an asymmetric offset split-gate voltage in the system described here will reduce the duct and so reduce the exchange contribution. Figure 4 in the previous section shows the results of such an experiment. The 0.25 plateau which we suggest is spin degenerate increases in magnitude with split-gate offset voltage and saturates at n = 1 confirming that the holes are now acting as independent particles with both spin directions and give the non-interacting value of conductance at $2e^2/h$. The 1/32 plateau which we suggest has a charge of e/4 and is spin polarised also increases in value but now saturates at n = 0.5 indicating that the holes are now acting as individual particles but are spin polarised as implied by the results discussed earlier with conductance at $e^2/h$.

**6. Summary**

The hole-hole interaction strongly modifies the sequence of quantum levels in the s-Ge quantum wire created by the spatial confinement. When the confinement is relaxed so that the system can expand in the second dimension there is a disappearance of the ground state, (that has half sine character), and it is replacement by a more complex structure producing fractional quantisation with charges e/2 and e/4. This observation suggests a new area of experimentation in quasi one-dimensional systems, particularly direct measurement of the charge, with implications for possible schemes of topological quantum information processing [34]. The clarity of the new states observed in s-Ge will also provide enhanced understanding to conductance structure below the first ballistic plateau found with electrons in GaAs.

*Acknowledgements*

This research was funded by the EPSRC with grant number EP/K004077/1 at the LCN and grant number EP/J003263/1 at Warwick University. Y. Gul acknowledges a CASE studentship with Toshiba Research Europe Limited.

References

[1] Myronov Maksym, Dobbie Andy, Shah Vishal A, Liu Xue-Chao, Nguyen van H and Leadley David R 2010 Electrochem. Solid State Lett. **13** H388; Myronov Maksym, Morrison Christopher, Halpin John, Rhead Stephen, Casteleiro Catarina, Foronda Jamie, Shah Vishal Ajit and David Leadley 2014 Japan. J. Appl. Phys. **63** 04EH02

[2] Morrison C and Myronov M 2017 Appl. Phys. Lett. **111** 192103; Mironov O A et al 2014 Phys. Status Solidi C **11** 61-64

[3] Stormer H L, Schlesinger Z, Chang A, Tsui D C, Gossard A C and Wiegmann W 1983 Phys. Rev. Lett. **51** 126

[4] Morrison C, Wisniewski P, Rhead S D, Foronda J, Leadley D R and Myronov M 2014 Appl. Phys. Lett. **105** 182401

[5] Holmes S N, Newton P J, Llandro J, Mansell R, Barnes C H W, Morrison C and Myronov M 2016 J. Appl. Phys. **120** 085702

[6] Thornton T J, Pepper M, Ahmed H, Andrews D and Davies G J 1986 Phys. Rev. Lett. **56** 1198

[7] Berggren K-F, Thornton T J, Newson D J, and Pepper M 1986 Phys. Rev. Lett. **57** 1769

[8] Wharam D A, Thornton T J, Newbury R, Pepper M, Ahmed H, Frost J E, Hasko D G, Peacock D C, Ritchie D A and Jones G A C 1988 J. Phys. C **21** L209

[9] van Wees B J, van Houten H, Beenaker C W J, Williamson J G, Kouwenhoven L P, van der Marel D and Foxon C T 1988 Phys. Rev. Lett. **60** 848

[10] Thomas K J, Nichols J T, Simmons M Y, Pepper M, Mace D R and Ritchie D A 1996 Phys. Rev. Lett. **77** 135

[11] Sfigakis F, Ford C J B, Pepper M, Kataoka M, Ritchie D A and Simmons M Y 2008 Phys. Rev. Lett. **100** 026807

[12] Patel N K, Nicholls J T, Martin-Moreno L, Pepper M, Frost J E F, Ritchie D A and Jones G A C 1991 Phys. Rev. B **44** 13549

[13] Chen T-M, Pepper M, Farrer I, Ritchie D A and Jones 2013 Appl. Phys. Lett. **103** 093503

[14] Meyer Julia S and Matveev K A 2009 J. Phys.: Condens. Matter **21** 023203

[15] Mehta Abhijit C, Umrigar C J, Meyer Julia S and Baranger Harold U 2013 Phys. Rev. Lett. **110** 246802

[16] Klironomos A D, Meyer J S and Matveev K.A 2006 Europhysics Letters **74** 679

[17] Welander E, Yakimenko I.I, and Berggren K-F 2010 Phys. Rev. B **82**, 073307

[18] Hew W K, Thomas K J, Pepper M, Farrer I, Anderson D, Jones G A C, Ritchie D A 2009 Phys. Rev. Lett. **102** 056804

[19] Smith L W, Hew W K, Thomas K J, Pepper M, Farrer I, Anderson D, Jones G A C and Ritchie D A 2009 Phys. Rev. B **80** 041306

[20] Kumar Sanjeev, Thomas Kalarikad J, Smith Luke W, Pepper Michael, Creeth Graham L, Farrer Ian, Ritchie David, Jones Geraint and Griffiths Jonathan 2014 Phys. Rev. B **90** 201304; Bayat Abolfazi, Kumar Sanjeev, Pepper Michael and Bose Sougato 2017 Phys. Rev. B **96** 041116(R)

[21] Sheng-Chin Ho et al, to be published

[22] Gul Y, Holmes S N, Newton P J, Ellis D J P, Morrison C, Pepper M, Barnes C H W and Myronov M 2017 Appl. Phys. Lett. 111, 233512

[23] Hew W.K, Thomas K.J, Pepper M, Farrer I, Anderson D, Jones G.A.C, and Ritchie D.A 2008 Phys. Rev. Lett. 101, 036801

[24] Laughlin R B 1983 Phys. Rev. Lett. **50** 1395

[25] de-Picciotto R, Reznikov M, Heiblum M, Umansky V, Bunin G and Mahalu D 1997 Nature **389** 162

[26] Maciejko Joseph and Fiete Gregory A 2015 Nature Physics **11** 385

[27] Su W P, Schrieffer J R and Heeger A J 1979 Phys. Rev. Lett. **42** 1698

[28 Su W P, Schrieffer J R and Heeger A J 1980 Phys. Rev. B **22** 2099

[29] Starykh O A, Maslov D L, Häusler W, and Glazman L I 2000 *Interactions and Quantum Transport Properties of Lower Dimensional Systems* (Lectures Notes in Physics vol 544) ed T. Brandes (New York: Springer) p 37

[30] Chen T-M, Graham A C, Pepper M, Farrer I and Ritchie D A 2008 Appl. Phys. Lett. **93** 032102

[31] Wu Congjun, Bergman Doron, Balents Leon, and Das Sarma S 2007 Phys. Rev. Lett. **99** 070401

[32] Zhang Machi, Hung Hsian-hsuan, Zhang Chuanwei, and Congjun Wu 2011 Phys. Rev. A **83** 0235615

[33] Thouless D J 1965 Proc. Phys. Soc. **86** 893

[34] Dolev M, Heiblum M, Umansky V, Stern Ady and Mahalu D 2008 Nature **452** 829; Radu Iuliana P, Miller J B, Marcus C M, Kastner M A, Pfeiffer L N and West K W 2008 Science 1157560; Quantum Information: Fraction Stations https://www.nature.com/articles/nnano.2008.112